\documentclass[
superscriptaddress,
preprint,
showpacs,
amsmath,amssymb,
aps,
]{revtex4-1}

\usepackage{multirow}
\usepackage{graphicx}
\usepackage{dcolumn}
\usepackage{epstopdf}
\usepackage{bm}

\begin{document}

\preprint{}

\title{Superhigh thermoelectric figure of merit in silver halides AgCl and AgBr from first principles}

\author{Xiuxian Yang}
\affiliation{School of Opto-electronic Information Science and Technology, Yantai University, Yantai 264005, People's Republic of China}
\author{Zhenhong Dai}
\email{zhdai@ytu.edu.cn}
\affiliation{School of Opto-electronic Information Science and Technology, Yantai University, Yantai 264005, People's Republic of China}
\author{Yinchang Zhao}
\email{y.zhao@ytu.edu.cn}
\affiliation{School of Opto-electronic Information Science and Technology, Yantai University, Yantai 264005, People's Republic of China}
\author{Sheng Meng}
\email{smeng@iphy.ac.cn}
\affiliation{Beijing National Laboratory for Condensed Matter Physics and Institute of Physics, Chinese Academy of Sciences, Beijing, 100190, People's Republic of China}
\affiliation{Collaborative Innovation Center of Quantum Matter, Beijing 100084, People's Republic of China}


\date{\today}

\begin{abstract}
Searching for the high-performance thermoelectric (TE) materials has always been a long-held dream in the thermoelectricity field. Recently, it is found in experiments that the largest figure of merit ZT of $2.6$ can be reached in SnSe crystals at $923$ K and Cu$_2$Se sample at $850$ K, which arouses the enormous interest of seeking high-ZT materials. Based on first-principle calculations and Boltzman transport equation (BTE), we report in this letter that silver halides (AgCl and AgBr) in rocksalt structure have excellent TE performances. A superhigh ZT of about $7.0$ at mid-temperature ($\sim600$ K) is obtained in the $p$-type doped AgCl and AgBr crystals, which far exceeds the ZT values of all current bulk TE materials. This record-breaking ZT value is attributed to the ultralow intrinsic lattice thermal conductivity $\kappa_L$ (e.g. $\kappa_L\sim0.10$ and $0.09$ Wm$^{-1}$K$^{-1}$ for AgCl and AgBr at $600$ K, respectively). Our results may be a feat that could revolutionize the field of the heat energy conversion.
\end{abstract}

\maketitle


\section{\label{intro}Introduction}
The thermoelectric (TE) technology provides a simple and environmentally friendly solution for direct conversion from heat to electricity, which has drawn a good deal of attentions\cite{Heeaak9997,111,PhysRevB.95.014307,ADMA:ADMA201602013,Bell1457,ADFM:ADFM200901905,PhysRevLett.88.216801}. However, due to lacking significant progress in lead-free, efficient TE devices, the influence of TE technology has mainly remained within a small sphere of niche applications\cite{10.1038/nature13184,C7EE01193H}. The efficiency of a TE material is determined by the figure of merit ZT defined as
\begin{equation}
ZT=\frac{S^2\sigma T}{\kappa_e+\kappa_L},
\end{equation}
where S is the Seebeck coefficient, $\sigma$ is the electrical conductivity, T is the temperature in kelvin, $\kappa_{e}$ represents thermal conductivity of the charge carrier and $\kappa_L$ is that of the lattice. Currently, the ZT values of most typical bulk TE materials, such as PbTe, $n$-Type skutterudites CoSb$_3$, PbSe, Bi$_2$Te$_3$, nanocomposites and $p$-type half-Heusler are in the range of $1.5\sim2.5$\cite{FU2016141,ROGL201430,Wang9705,10.1038/ncomms10766,10.1038/ncomms13713,10.1038/ncomms9144,Kim109}. Due to the interdependent relationship in $\sigma$, S and $\kappa$, it is usually difficult to improve an average ZT well above $2.5$. In the past decade, there are several approaches proposed to improve ZT, including the enhancement of electronic properties (S and $\sigma$) (by doping electron or hole\cite{zebarjadi2011power,yu2012enhancement}, introducing the resonant states in the vicinity of Fermi level\cite{heremans2012resonant,heremans2008enhancement,minnich2009bulk}, and band convergence\cite{liu2012convergence,liu2013low}) and the reduction of the $\kappa_L$ (by enhancing phonon scattering through disorder within the unit cell\cite{markussen2009electron,bhattacharya2006effect} or forming solid solutions\cite{liu2012convergence,poudeu2006nanostructures}).

So far, the largest ZT value of $2.6$ in experiment can be reached in SnSe crystals at $923$ K and Cu$_2$Se sample at $850$ K\cite{10.1038/nature13184,C7EE01193H}. In SnSe crystals the excellent TE performance stems from the ultralow thermal conductivity, such as $\kappa_L\sim0.23\pm0.03$ Wm$^{-1}$K$^{-1}$ at $973$ K, while the remarkable TE performance in Cu$_2$Se sample is attributed to the localization of Cu$^+$ induced by the incorporation of indium (In) into the Cu$_2$Se lattice, which enhances the $\sigma$ and reduces the $\kappa_L$ of the nanocomposites simultaneously\cite{10.1038/nature13184,C7EE01193H}. A common feature in these materials is the presence of low $\kappa_L$, which is a crucial ingredient of high-ZT materials. Therefore, searching for TE materials with low intrinsic $\kappa_L$ is our striving directions. In $1986$, M. V. Smirnov $et$ $al.$ reported that molten alkali halides and their mixtures have ultralow $\kappa_L$, e.g., the $\kappa_L\sim0.2-1.4$ W/mK in the temperature range of $900\sim1300$ K\cite{smirnov1987thermal}, which inspires us to study the heat transport and TE properties of halide materials. In this work, we present that the silver halides AgCl and AgBr crystals materials may be the best candidates for high TE performance.

For AgCl and AgBr, the theoretical and experimental studies are only focused on the electronic structure properties, ionic transport properties, optical absorption and response or other chemical properties in the past few decades\cite{doi:10.1063/1.473690,PhysRevB.11.1654,PhysRev.97.676,joesten1966indirect,carrera1971optical,PhysRev.137.A1217}. In the practical applications, AgCl is a common reference electrode in electrochemistry, while AgBr is widely used in photographic films. Nevertheless, to date the study of the heat transport properties and TE properties in these materials is lacking, which is may be due to the low melting point ($728$ K for AgCl, $701$ K for AgBr)\cite{PhysRevB.11.1654,doi:10.1063/1.473690}. In this paper, we systematically investigate the heat and electronic transport properties of AgCl and AgBr, and conclude that they have ultralow intrinsic $\kappa_L$, high S, and consequently remarkable TE performances.

\section{\label{method}Methodology}

Using first-principle calculations and Boltzman transport equation (BTE), we study the electronic structure, lattice thermal transport and electron transport properties of AgCl and AgBr crystals with rocksalt structure. The calculations are performed by the Vienna Ab-initio Simulation Package (VASP)\cite{KRESSE199615,PhysRevB.78.134106}, which is based on the density functional theory (DFT). In the DFT calculations, a $520$ eV energy cutoff with the exchange-correlation functional of generalized gradient approximation (GGA) of the Perdew-Burke-Ernzerhof (PBE)\cite{PhysRevLett.77.3865} is used to simulate the valence electron. A $25\times25\times25$ k-point is utilized for the electron-momentum integration. We use the ShengBTE\cite{17li2014shengbte} package to calculate the lattice thermal conductivity with a $30\times30\times30$ q-mesh. The only input parameters are the harmonic and anharmonic interatomic force constants (IFCs). The harmonic IFCs were obtained based on the finite-difference approach via the PHONOPY program\cite{togo2008first} within the $5\times5\times5$ supercells, and the anharmonic IFCs were created by thirdorder.py script\cite{17li2014shengbte} within the $5\times5\times5$ supercells. In the anharmonic IFCs calculations, the eight nearest neighbor interactions were taken into account.

To obtain the electron transport properties, we use the rigid-band approach and the semiclassical Boltzmann theory, which is performed in the BOLTZTRAP code\cite{MADSEN200667}. In this approach, the constant scattering time approximation $\tau$ is used, which is the only parameter that can be tuned. The value of $\tau$ from $1$ to $8$ fs is used to obtain relatively reasonable results. The electronic structure is recalculated by VASP on the dense k-points of $80\times80\times80$ to acquire precise derivatives of the Kohn-Sham eigenvalues.

\section{\label{result}Results and discussion}

\begin{table*}[]
\caption{\label{tab:table1} The lattice constant a ({\AA}), volume V({\AA}$^3$/(unit cell)), comparison of bang-gap (eV) obtained in GGA and GW theoretical calculations compared with experimental values, the melting point of AgCl and AgBr crystals.}
\begin{ruledtabular}
\begin{tabular}{ccccccc}
  \multicolumn{1}{c}{Crystal}  &\multicolumn{1}{c}{a}    &\multicolumn{1}{c}{V}    &\multicolumn{1}{c}{E$_{gap}^{GGA}$}  &\multicolumn{1}{c}{E$_{gap}^{GW}$} &\multicolumn{1}{c}{E$_{gap}$(Ex)} &\multicolumn{1}{c}{melting point} \\
  \multicolumn{1}{c}{}  &\multicolumn{1}{c}{${\AA}$} &\multicolumn{1}{c}{${\AA}^3$/unit cell} &\multicolumn{1}{c}{eV} &\multicolumn{1}{c}{eV} &\multicolumn{1}{c}{eV} &\multicolumn{1}{c}{K} \\
\hline \\
  AgCl      &5.60 &44.02 &0.95 &3.28  &3.0(Ref.\cite{PhysRevB.56.4417}) &728(Ref.\cite{PhysRevB.11.1654,doi:10.1063/1.473690})\\ \\
  AgBr      &5.85 &49.82 &0.70 &2.70  &2.5(Ref.\cite{PhysRevB.56.4417}) &701(Ref.\cite{PhysRevB.11.1654,doi:10.1063/1.473690}) \\
\end{tabular}
\end{ruledtabular}
\end{table*}

\begin{figure}
\includegraphics[width=8 cm]{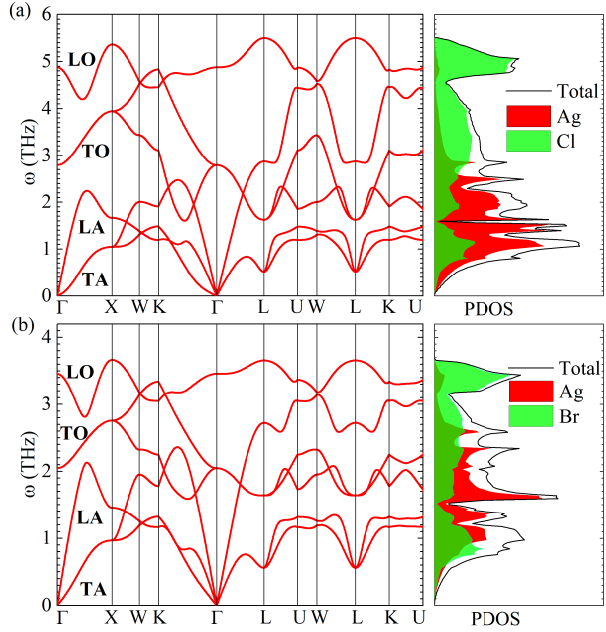}
\caption{(Color online). Calculating phonon dispersion relations and phonon density of states (PDOS) of AgCl (a) and AgBr (b) along the high symmetry point within the first Brillouin zone (BZ), respectively.}
\end{figure}

AgCl and AgBr crystals are stable in the rocksalt structures (as shown in Supplemental S.~$1$) below their melting point ($728$ K for AgCl, $701$ K for AgBr)\cite{PhysRevB.11.1654,doi:10.1063/1.473690}. Their optimized lattice constants $a$ are $5.60$ {\AA} and $5.85$ {\AA}, respectively, as shown in TABLE I, which are slightly larger than the experimental values of $5.55$ {\AA} for AgCl and $5.774$ {\AA} for AgBr\cite{PhysRev.97.676}, this is due to the fact that GGA often overestimates the lattice constant. Usually, the phonon spectrum can inspect the structure stability, which lies in the fact that for each phonon mode the frequency should be a real quantity and not imaginary\cite{TOGO20151}. Figure~$1$ illustrates the phonon dispersion spectrum and phonon density of states (PDOS) of AgCl and AgBr. Obviously, there are no imaginary frequencies in the phonon spectrum, indicating that these crystals are stable structures. In our calculations, phonon spectrum of AgCl and AgBr are consistent with other theoretical calculation and experimental data\cite{PhysRevB.74.054102,vaidya1971compressibility,dorner1976w}, which indicates the accuracy of our calculations. There are six phonon modes: two transverse acoustic modes (TA), one longitudinal acoustic mode (LA), two transverse optic modes (TO) and one longitudinal optic mode (LO) in the phonon spectrum due to the existence of two atoms in the unit cell. The optic modes exhibit large splitting of TO and LO modes around the $\Gamma$ point because of the strong coupling between the lattice and polarization filed. The polarization filed is induced by the longitudinal optic modes in the phonon long-wavelength limit in the ionic crystals. The polarization field depends on the dielectric constants and the Born effective charges computed by the density functional perturbation theory (DFPT) in the VASP code. For AgCl (AgBr), the Born effective charges of Ag atom and halogen atom are $1.45$ ($1.53$) $e$ and $-1.45$ ($-1.53$) $e$, respectively, and the dielectric constant is $4.94$ ($5.87$), indicating a strong polarization field and thus resulting in a large TO/LO splitting of about $2.13$ ($1.42$) THz, as presented in Fig.~$1$. The partial PDOS shows that for AgCl the acoustic modes are mainly afforded by the Ag atom, while for AgBr the acoustic modes are afforded by the combination of Ag atom and Br atom. This is attributed to the difference of the atomic effective mass.

\begin{figure}
\includegraphics[width=8 cm]{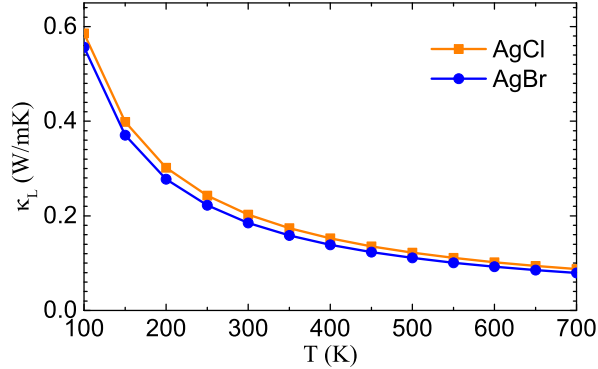}
\caption{(Color online). The lattice thermal conductivity $\kappa_L$ versus temperature T for AgCl and AgBr. The origin square and blue circle lines represent the $\kappa_L$ for AgCl and AgBr, respectively.}
\end{figure}

Figure~$2$ gives intrinsic $\kappa_L$ of AgCl and AgBr versus temperature from $100$ to $700$ K. These silver halide materials have smaller discrepancy of $\kappa_L$ and exhibit similar $\kappa_L\varpropto T^{-1}$ trend. Since more phonons are active under high temperature, the Umklapp process will become critical in the phonon scattering and reduce the $\kappa_L$\cite{DING20171164}. Remarkable, the values of $\kappa_L$ are fairly low. For instance, the $\kappa_L$ of $0.202$ ($0.102$) and $0.185$ ($0.093$) Wm$^{-1}$K$^{-1}$ are obtained at $300$ ($600$) K in AgCl and AgBr crystals, respectively. These values of $\kappa_L$ are much lower than the commercial TE materials PbTe ($1.40-2.85$ Wm$^{-1}$K$^{-1}$ at $300$ K) and its alloys ($1.78$ and $1.42$ Wm$^{-1}$K$^{-1}$ for Pb$_{0.94}$Mg$_{0.06}$Te and Pb$_{0.8}$Mg$_{0.2}$Te at $300$ K, respectively.)\cite{FU2016141}. Furthermore, to gain insight into the heat transport mechanism and decide which phonon modes provide the primary heat conductivity, the accumulative lattice thermal conductivity $\kappa_a$ scaled by the total $\kappa_L$ as a function of frequency, which exhibits the summed contributions from the phonon modes below the specified frequency, is also calculated, as shown in Supplemental S.$2$. For AgCl (AgBr), more than $80\%$ of the heat transport is induced by the phonons with the frequency below $3$ ($2$) THz. Based on the combination of S.$2$ and Fig.~$1$, we can find that three acoustic phonon modes and two TO modes dominate the heat transport. In addition, the ultralow $\kappa_L$ usually hints excellent TE performance if their power factor $S^2\sigma$ (PF) and electronic transport properties are good enough.

\begin{figure}
\includegraphics[width=8 cm]{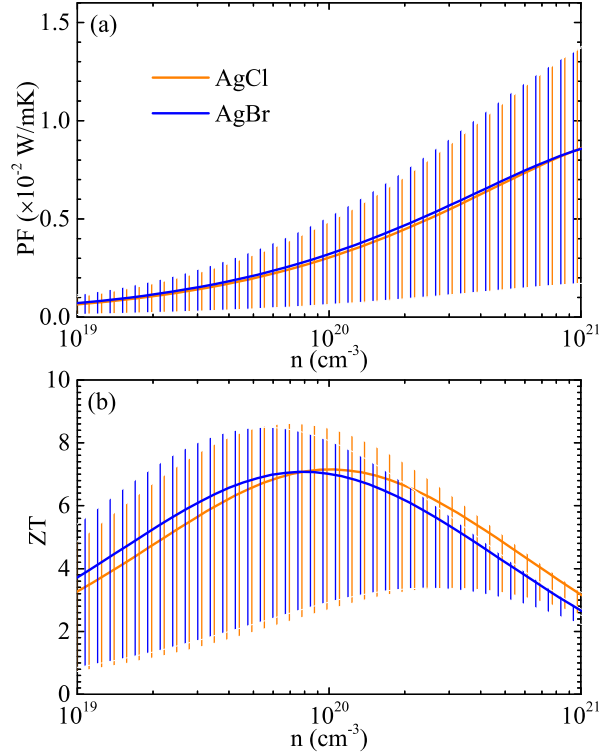}
\caption{(Color online).  At $600$ K, the TE parameters: (a) the power factor (PF) $S^2\sigma$ and (b) figure of merit ZT for the p-type doped AgCl (orange curves) and AgBr (blue curves) as a function of carrier concentration. In these panels, the lower and upper limits of the vertical bars and the full curves represent the values with $\tau=1, 8$ and $5$ fs, respectively.}
\end{figure}

The electronic structure diagrams of AgCl and AgBr are shown in Supplemental S.~$3$(a-b), which are obtained from GGA method. The GGA results show that AgCl and AgBr are indirect band gap semiconductors with the band gap values of $0.95$ and $0.70$ eV. The conduction band minima (CBM) locates at the high-symmetry L point and valence band maxima (VBM) locates at the high-symmetry $\Gamma$ point. It must be point out that the GGA results are much lower than the experimental values (such as, $3.0$ eV for AgCl and $2.5$ eV for AgBr\cite{PhysRevB.56.4417}), as shown in TABLE I. Therefore, we used the GW$_0$ method\cite{PhysRevB.75.235102,PhysRevB.76.115109,PhysRevB.74.035101,PhysRevLett.99.246403} to recalculate the electronic band structure. These results are shown in Supplemental S.~$3$(c-d) and TABLE I. The GW$_0$ results reveal that AgCl and AgBr are also indirect band gap semiconductors but with the wide band gap of $3.28$ and $2.70$ eV, which are consistent with the experimental values. Since the electronic transport properties are determined by the electronic energy band structure, the project band structure diagrams and partial electronic density of states (EDOS) for AgCl and AgBr calculated with GGA are shown in Supplemental S.~$4$(a-b), respectively. The partial EDOS shows that the valence band close to Fermi level is mainly contributed by the Ag atom d orbital and halogen atom p orbital. The dispersion of energy band structure reveals both heavy and light effective masses of charge carriers. The flat band is in the valence band along $\Gamma$-K line, making a large EDOS and heavy effective masses, which leads to high thermal power S. In contrast, the strong dispersion are observed in both valence band along K-$\Gamma$-L line and conduction band minimum, which indicates a high electron mobility, $\mu$. Thus, according to the equation of $\sigma=nq\mu$, here $n$ is the carrier concentration, $q$ is quantity of electric charge and $\sigma$ is the electrical conductivity. Thus, a high $\sigma$ can be expected, if $n$ is high enough.

To evaluate the electronic transport and TE properties, the S, $\sigma/\tau$ and $\kappa_e$ were calculated by the BOLTZTRAP code\cite{MADSEN200667} as a function of temperature and carrier concentration $n$. The calculations with the scissors shift of $2.33$ and $2.0$ eV (E$_{gap}^{GW_0}$-E$_{gap}^{GGA}$) for AgCl and AgBr crystals, are also performed. The results show that band gap underestimate have no effect on electronic transport properties. Additionally, the best TE performances are obtained in $p$-type doped AgCl and AgBr crystals, while $n$-type doping cases show bad TE performance, as shown in Supplemental S.~$5$, thus we no longer care about the $n$-type doping cases for these materials in the following discussion.

The calculated thermal power S is shown in Supplemental S.~$6$. The S increases with temperature at the same carrier concentration and decreases with carrier concentration at the same temperature, similar to the tendency in most of semiconducting TE materials\cite{PhysRevB.95.014307}. The S values are much high. For instance, the values of S for AgCl (AgBr) are in the range of $350\sim430$ ($330\sim410$) $\mu V/K$ at n$\sim10^{20}$ $cm^{-1}$ as the temperature increases from $300$ to $600$ K. These values of S are much larger than PbTe, such as the values of S are in the range of $100\sim300$ $\mu$V/K at $300\sim600$ K\cite{Heremans554}. In addition, the values of S of AgCl are higher than that of AgBr at the same temperature, indicating that a possible higher ZT in AgCl. Although the $\sigma$ and $\kappa_e$ values are not confirmed at present, the combination of large S and ultralow $\kappa_L$ suggest a possible high ZT in AgCl and AgBr. To calculate $\sigma$ and $\kappa_e$, we should estimate the amplitude of electronic scattering times $\tau$ and consider the effect of lattice vibration on practical $\tau$, thus we use $\tau$ from $1$ to $8$ fs to obtain a reasonable result, as shown in Supplemental S.~$7$ and S.~$8$. We find that the $\sigma$ values does not rely on temperature, which is in accordance with the electronic Boltzmann theory\cite{MADSEN200667}. The values of $\sigma$ of AgCl are slightly smaller than that of AgBr. Moreover, to obtain the accuracy ZT values, it is necessary to calculate the $\kappa_e$, although the values of $\kappa_e$ are much lower than the $\kappa_L$. These results are presented in Supplemental S.~$8$ at $300, 500$ and $600$ K for AgCl and AgBr, respectively. One can find that the values of $\kappa_e$ of AgBr are slightly higher than that of AgCl in the temperature range $300\sim600$ K.

Next, the power factor (PF) and ZT for p-type doped AgCl and AgBr as a function of carrier concentration at $600$ K are shown in Fig.~$3$(a) and (b). In these panels, the lower and upper limits of the vertical bars and the full curves represent the values with $\tau=1, 8$ and $5$ fs, respectively. The PF curves of AgBr are slightly higher than that of AgCl mainly due to the higher $\sigma$ in AgBr crystal. The values of ZT with $\tau=1\sim8$ fs in the wide carrier concentration region (n$=10^{17}\sim10^{22}$ cm$^{-1}$) at $300$, $500$ and $600$ K for AgCl and AgBr are presented in Supplemental S.~$9$(a-c). For AgCl, with $\tau=8$ fs, the extraordinarily high ZT of $8.62$ is obtained at $600$ K and n$\approx7.32\times10^{19}$ cm$^{-1}$ when PF$\sim0.41$ Wm$^{-1}$K$^{-1}$. As $\tau$ decreases to $5$ fs, a fairly large ZT is $7.23$ in higher n ($\sim1\times10^{20}$ cm$^{-1}$) when PF$\sim0.32$ Wm$^{-1}$K$^{-1}$. Even with $\tau=1$ fs, a large ZT ($3.44$) is also achieved in n$\sim3.24\times10^{20}$ cm$^{-1}$ when PF$\sim0.11$ Wm$^{-1}$K$^{-1}$. For AgBr, with $\tau=8$ fs, the extraordinarily high ZT of $8.46$ is obtained at $600$ K and n$\approx5.36\times10^{19}$ cm$^{-1}$ when PF$\sim0.35$ Wm$^{-1}$K$^{-1}$. A fairly large ZT of $7.01$ is obtained in n$\approx8.12\times10^{19}$ cm$^{-1}$ as $\tau$ decreases to $5$ fs when PF is $0.29$ Wm$^{-1}$K$^{-1}$. As $\tau$ further decreases to $1$ fs, ZT of $3.49$ is also larger with higher n ($\sim2.49\times10^{20}$ cm$^{-1}$) when PF is $0.11$ Wm$^{-1}$K$^{-1}$. It should be noted that the maximum ZT need lower carrier concentration as $\tau$ increases. At the lower temperature, such as $300$ K, AgCl and AgBr also exhibit excellent TE performance. The high ZT values are $3.09$ and $2.39$ for AgCl when $\tau$ is $8$ and $5$ fs, respectively, as shown in Supplemental S.~$9$(a) the real curves. For AgBr, the ZT values are $2.89$ and $2.25$ when $\tau$ is $8$ and $5$ fs, respectively, as shown in Supplemental S.~$9$(a) the dash curves. At $500$ K, for AgCl (AgBr) the ZT values are $6.83$ ($6.72$) and $5.60$ ($5.49$) when $\tau$ is $8$ and $5$ fs, respectively, as shown in Supplemental S.~$9$(b). One can find that large PF can lead to a high ZT, meanwhile being subject to $\kappa_e$ and $\kappa_L$, which indicates a compromise between the PF and $\kappa$ and a complex competition mechanism within the TE materials. Our results suggest that AgCl and AgBr crystals in rocksalt structure have unprecedented large ZT values, which are highest than that of all current bulk TE materials. Finally, we must point out that due to the property of unusual sensitivity to light for these materials, a black shell is needed to design TE devices, meanwhile ensuring that the operating temperature of devices is lower than their melting point.

\section{conclusion}

To summarize, we have calculated electronic structure, lattice thermal transport and electronic transport properties of rocksalt structure AgCl and AgBr crystals, which is employed first principles and phonon (electron) Boltzmann transport theory. The ultralow $\kappa_L$ of $0.202$ and $0.185$ Wm$^{-1}$K$^{-1}$ of AgCl and AgBr are obtained at the room temperature. Usually, the ultralow $\kappa_L$ indicates the excellent TE performances, hence a combination of the first principle calculations and the semiclassical analysis was used to investigated the TE properties for these materials. The electronic transport properties are determined by the electronic energy band structure. We find that the flat band leads to high S and highly dispersive band results in good $\sigma$. Therefore, the unprecedentedly large values of ZT of $7.2$ and $7.1$ are obtained at $600$ K in the p-type doped AgCl and AgBr, which is defeated the ZT values of all current bulk TE materials. These results indicate that AgCl and AgBr are excellent mid-temperature ($500-900$ K) power generation materials, although we need a black shell and the operating temperature of devices is below their melting point.

\section{acknowledgment}
This research were supported by the National Natural Science Foundation of China under Grant No.11774396 and No.11704322, Shandong Natural Science Funds for Doctoral Program under Grant No.ZR2017BA017, the National Key Research and Development Program of China under Grant No.2016YFA0300902, and Graduate Innovation Foundation of Yantai University, GIFYTU, No.YDZD1810.

\nocite{*}

\bibliography{apsref}

\end{document}